\documentclass[conference]{IEEEtran}
\IEEEoverridecommandlockouts
\usepackage{cite}
\usepackage{amsmath,amssymb,amsfonts}
\usepackage{algorithmic}
\usepackage{graphicx}
\usepackage{textcomp}
\usepackage{xcolor}

\usepackage{multirow}
\usepackage{array}
\newcolumntype{?}{!{\vrule width 0.5pt}}
\usepackage{makecell}
\usepackage{nidanfloat} 

\usepackage{booktabs}
\usepackage{tabto}

\usepackage{url}

\usepackage{flushend}

\def\BibTeX{{\rm B\kern-.05em{\sc i\kern-.025em b}\kern-.08em
    T\kern-.1667em\lower.7ex\hbox{E}\kern-.125emX}}
\begin{document}

\title{Predicting Relative Thresholds\\for Object Oriented Metrics}

\author{\IEEEauthorblockN{Sultan Alhusain}
\IEEEauthorblockA{\textit{College of Computing and Informatics} \\
\textit{Saudi Electronic University}\\
Riyadh, Saudi Arabia \\
s.alhusain@seu.edu.sa}
}

\maketitle

\begin{abstract}
Object-oriented software metrics provide a numerical characterization of software quality. They have also been used in the assessment and identification of technical debt. However, metrics generally need to be used with thresholds as reference points that help to interpret their values properly and objectively. The problem is that, while there are many proposed metrics, there are relatively few studies on thresholds and threshold calculation methods; hence, the effective application of metrics in practice has been limited. Moreover, although it has been acknowledged that thresholds should not be absolute, but rather relative to certain contextual factors, the context is still not considered in most threshold studies. In~this~paper, the relationship between system size (as a contextual factor) and metric thresholds is investigated. The objective is to build predictive models that estimate thresholds based solely on system size, and to assess the feasibility of this approach as a threshold estimation method. An empirical study is conducted for this purpose using $36$ defect-prediction datasets and six metrics. The results show that the proposed threshold estimation method is feasible, and it can achieve an accuracy remarkably comparable to more complex threshold models.
\end{abstract}


\section{Introduction}

Many object-oriented (OO) metrics have been proposed and empirically validated to facilitate the production of quality software products \cite{110} \cite{105}. They typically provide a numerical characterization of an internal quality attribute (e.g. coupling) that is associated with some external quality attribute(s) (e.g. fault-proneness) \cite{106}. The existence of such associations can be explained by the hypothesis proposed in \cite{107}, which states that the structural complexity of a software system mirrors the cognitive complexity required to deal with it. Consequently, the increase in cognitive burden on system developers increases the likelihood of producing undesirable external quality attributes.

Software metrics generally serve two primary purposes: the construction of classification models to identify faulty classes, and the formulation of guidelines to prevent the introduction of faults in the first place \cite{106}. They have also been used for the purpose of assessing and identifying technical debt \cite{301} \cite{300} \cite{299}. These purposes are often better served by using thresholds, which provides an easy and appealing method for applying metrics in practice. Thresholds can be defined as the values below which risks are quantifiable and acceptable \cite{114}. Each threshold represents a reference point with which metric values can be objectively interpreted. These reference points can be used to inform design decisions or to optimally allocate testing resources \cite{109}.


While there are many proposed metrics, relatively little research has been done to identify thresholds for them \cite{287}. Moreover, most of the published work on thresholds does not take contextual variables into account \cite{288}, although it has been acknowledged that thresholds should not be \textsl{absolute} but rather \textsl{relative} to certain contextual factors \cite{122}. In fact, the distribution of metric values can vary between contexts \cite{117}, and such variations affect what are (and are not) reasonable and useful thresholds. An exceedingly high metric value in one context may be relatively low in another. The effect of these variations in distribution may not be limited to thresholds, as more complex models can be affected as well \cite{290}. Thus, there are still no universally accepted thresholds, even for the most studied metrics, and the effective application of metrics in practice has so far been limited \cite{116} \cite{113} \cite{289}.






In this study, the objective is to build predictive models that estimate thresholds based solely on system size, and to assess the feasibility of this approach as a threshold estimation method. To this end, thresholds are calculated for six metrics in $36$ software systems of various sizes. The size--threshold relationships are then examined by analyzing the correlation between the two for each metric. If such a relationship exists for a metric, a linear regression model is fitted and used to estimate its thresholds given only system sizes. The feasibility of this approach is then assessed based on the accuracy with which the estimated thresholds identify fault-prone classes. This assessment strategy is supported by the fact that producing good reliability predictors is a major objective for software metrics research \cite{159}.


The thresholds used in the analysis of the size--threshold relationship are calculated using the logistic regression (LR) threshold model, which was initially introduced in the field of epidemiology \cite{114} and later used in several software engineering studies \cite{112} \cite{294}. However, an improvement to this LR threshold model is proposed and used in this study. The improved model aims to address the problem of generating negative thresholds in some cases, as reported in the literature (e.g., \cite{291}), and to counteract the effect of data imbalance on the LR model.

The selection of size as a contextual factor is motivated by what could be described as a trend of positive relationships between system size (as measured by the number of classes) and thresholds calculated using different calculation methods (Table~\ref{tab:ThresholdsByOthers}). An observation of a similar trend is also reported in \cite{293}, although the size in their study is measured by lines of code. Moreover, it is suggested in \cite{291} that thresholds derived from small systems may not be useful in larger ones. The selection of this contextual factor is also motivated by the fact that system size can easily be incorporated into mathematical models, which can be used to estimate thresholds given only system size as input.

\begin{table*}[!b]
\renewcommand{\arraystretch}{1.65}
  \centering
  \caption{Thresholds as Reported in the Literature for Small (S), Medium (M), Large (L) and Larger (Lx) Systems}
	\begin{minipage}{\textwidth}
	\centering
    \begin{tabular}{?c?c|c|c?c|c?c|c|c|c?}
		\Xhline{0.5pt}
    \multirow{2}[4]{*}{\textbf{Metric Name}} & \multicolumn{3}{c?}{\cite{111}\textsuperscript{$\ast$}} & \multicolumn{2}{c?}{\cite{108}} & \multicolumn{4}{c?}{\cite{126}} \\
		\cline{2-10}
          & \textbf{S} & \textbf{M} & \textbf{L} & \textbf{S} & \textbf{L} & \textbf{S} & \textbf{M} & \textbf{L} & \textbf{Lx} \\
		\Xhline{0.5pt}
    Coupling Between Object classes (CBO) & 8     & 13    & 10    & 1     & 8     & 4     & 8     & 17    & 25 \\
		\Xhline{0.25pt}
    Response For a Class (RFC) & 36    & 39    & 44    & 27    & 25    & 17    & 5     & 24    & 48 \\
		\Xhline{0.25pt}
    Coupling Through Message passing (CTM) & 30    & 33    & 35    &       &       &       &       &       &  \\
		\Xhline{0.25pt}
    Weighted Methods per Class (WMC) & 21    & 18    & 24    & 11    & 31    & 18    & 46    & 17    & 43 \\
		\Xhline{0.25pt}
    Number of Methods (NOM) & 8     & 8     & 9     &       &       & 1     & 3     & 11    & 32 \\
		\Xhline{0.5pt}
		\multicolumn{10}{c}{\textsuperscript{$\ast$}Size information is taken from \cite{137}.}
    \end{tabular}%
  \label{tab:ThresholdsByOthers}%
\end{minipage}
\end{table*}%

In fact, because the majority of OO metrics have been found to follow a power law distribution \cite{160} \cite{120}, the largest value of these metrics increases with the number of classes in a software system. Chidamber and Kemerer noted this effect and stated as a possible explanation that ``coupling between classes is an increasing function of the number of classes in the application'' \cite[page 487]{121}. This explanation is consistent with the results of the study reported in \cite{119}, which also suggests that some high coupling is impractical to eliminate, and that it is even necessary for good design.

The remainder of this paper is organized as follows. The improvement proposed to the LR threshold model is presented and discussed in the subsequent section. In Section~\ref{sec:Method}, the research method used in this study is described. The results and threats to validity are then presented and discussed in Sections~\ref{sec:Results} and ~\ref{sec:ValidityThreats}, respectively. Related work is discussed in Section~\ref{sec:Related}. The conclusions and outlines of future work are finally discussed in Section~\ref{sec:Conclusion}.


 \section{Improved LR Threshold Model}
\label{sec:Effect}

To examine the size–threshold relationship, and subsequently build the threshold estimation models, we first need to calculate thresholds for different metrics in different software systems of various sizes. This is done using the LR threshold model \cite{112}. While the original LR model estimates the probability of having a defect in a class based on a given metric value, the LR threshold model calculates a value (threshold) based on a given acceptable risk (probability) level. However, there are two problems with the LR threshold model. The first problem is the generation of negative thresholds for some risk levels (as reported, for example, by \cite{112}, \cite{294} and \cite{291}). After examining this issue, the cause for the negative thresholds has been determined to be the setting of acceptable risk levels that are below the background risk level (i.e. the risk corresponding to the metric value of zero).


The second problem is caused by using imbalanced datasets, and this problem is often ignored in software engineering studies \cite{130}. This imbalance affects one of the coefficients of the LR model, and it leads to an underestimation of the risk (probability) for the category with fewer samples, which consequently affect the calculated threshold values. Addressing this issue is particularly important when dealing with defect prediction datasets as many, if not most, of them are imbalanced. The solutions to both problems are presented as follows.
\vspace{+1mm}

\noindent The general model of LR is:
\begin{equation}
P(x) = \frac{e^{\alpha + \beta x }}{1 + e^{\alpha + \beta x }}
\end{equation}
where $P(x)$ is the probability (risk) of having a defect, given $x$ as a metric value, and $\alpha$ and $\beta$ are the parameters to be estimated. The risk when a metric value equals zero ($p_0$) can be calculated by the following equation:
\begin{equation}
p_0 = P(0) = \frac{e^{\alpha }}{1 + e^{\alpha}}
\label{eq:probAtZero}
\end{equation}
The risk denoted by $p_0$ is the background risk. The solution to the first problem lies in not setting the Acceptable Risk Level ($ARL$) arbitrarily but, rather, systematically by considering the background risk. To do so, the $ARL$ can be calculated using an acceptable increase to the background risk. If the acceptable increase is set to $0.10$, for example, the $ARL$ can simply be calculated as:
\begin{equation}
P_{ARL} = p_0 + 0.10
\label{eq:ch4Parl}
\end{equation}
Thresholds (i.e. metric Values at the ARL) can then be calculated using the equation:
\begin{equation}
VARL = \frac{1}{\beta}(\ln(\frac{P_{ARL}}{1- P_{ARL}})-\alpha)
\label{eq:VARL}
\end{equation}
It should be noted that the $P_{ARL}$ calculated in Eq.~(\ref{eq:ch4Parl}) must be $<1$ since Eq.~(\ref{eq:VARL}) is not defined for $P_{ARL}\geq 1$. 



\begin{figure}[!b]
\centering
\includegraphics[width=\columnwidth]{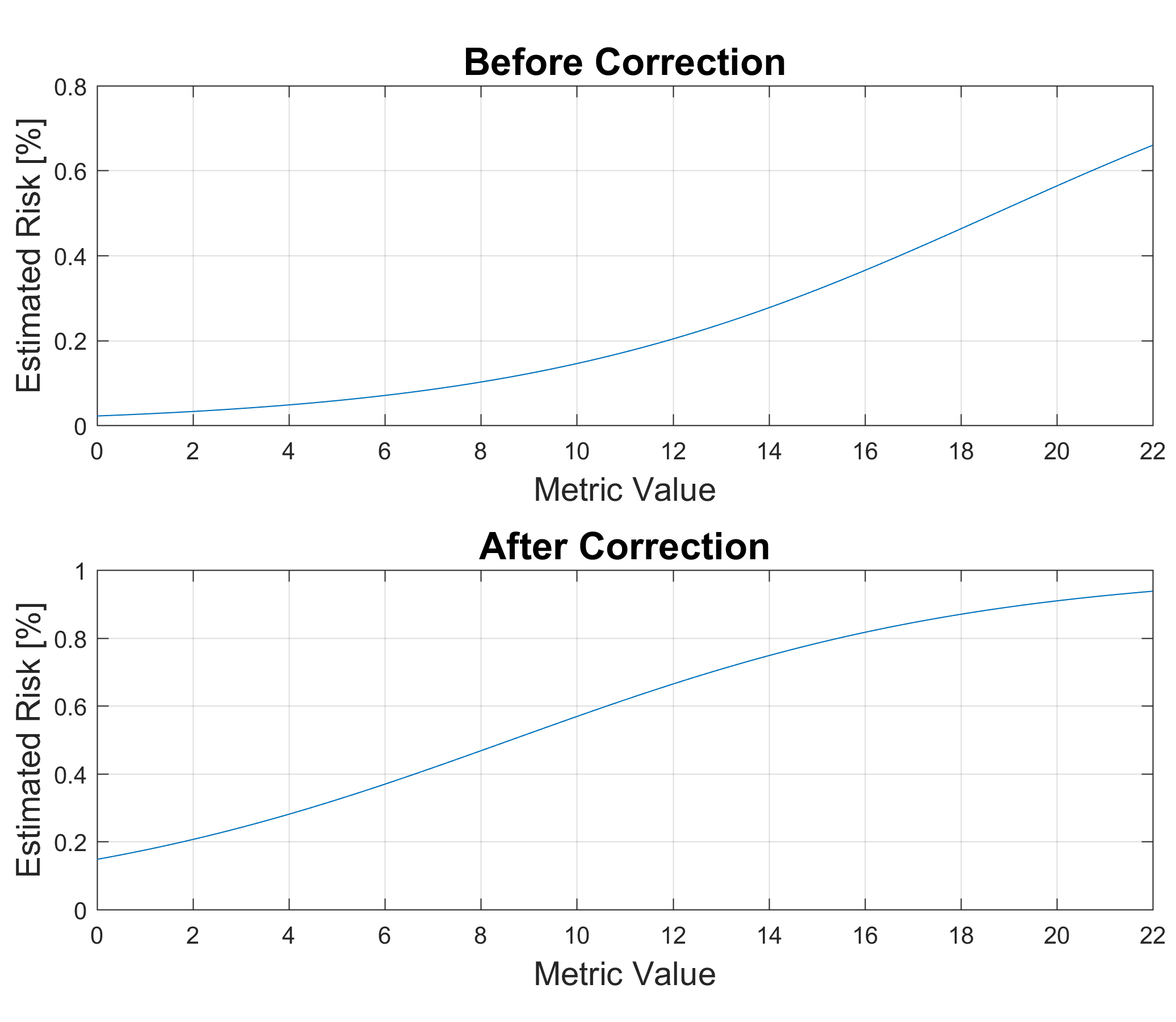}
\caption{An Example for the Estimated Risk Before and After the Correction}
\label{camelOneZero}
\end{figure}

The imbalance problem affects the intercept coefficient ($\alpha$), which may distort the background risk calculation in Eq.~(\ref{eq:probAtZero}) as well as the threshold calculation in Eq.~(\ref{eq:VARL}). Although such a distortion in $\alpha$ can be easily corrected using the correction proposed in \cite{136}, the correction needs information about the ratio of defects in the population ($y$) in addition to their ratio in the datasets of individual systems ($\bar{y}$). Thus, in order to be able to apply the correction, we would need to assume that the average ratio of defects in all of the systems included in this study is the ratio of defects in the population. The corrected value $\hat{\alpha}$ can then be calculated using the following equation \cite{136}:
\begin{equation}
\hat{\alpha} = \alpha - \ln((\frac{1-y}{y})(\frac{\bar{y}}{1-\bar{y}}))
\label{eq:ch4correction}
\end{equation}
Fig.~\ref{camelOneZero} shows an example of how this correction changes the estimated risk of having a defect across a range of metric values. For comparison, the results will be presented with and without the correction (by using the coefficients $\hat{\alpha}$ and $\alpha$, respectively), as will be discussed later.

\section{Methodology}
\label{sec:Method}

\subsection{Dataset Preparation}

A total of $36$ defect datasets are used in this study, representing different versions of $13$ open source Java systems. The size of the systems varies from $26$ to $1450$ classes. The average ratio of defects in all of the datasets is approximately $0.358$, which will be used in the correction equation as discussed above. The datasets are obtained from two publicly available repositories (PROMISE \cite{131} and Bug Prediction Dataset \cite{132}). Each dataset contains a list of classes along with the number of defects found in each class. However, the classes are re-labeled as faulty (i.e. number of defects $\geq 1$) or not-faulty because the LR model requires a binary dependent variable. Also, because the size of the systems is an important variable of interest, classes from external libraries are removed and not counted (e.g., the classes of the Regex regular expression library in the datasets of the JEdit system). Counting such classes when measuring system sizes may obscure or distort the size--threshold relationships, if any exist.


As noted in \cite{130}, there are inconsistencies between the datasets of different software repositories because different tools are used to construct them. Therefore, in order to make the datasets obtained from the two repositories consistent, they were all regenerated\footnote{All of the regenerated datasets are available online at:\tab \tab \url{https://zenodo.org/record/4625975}.} using a single tool. The tool is developed using the POM metric calculation framework \cite{295} as well as ASM bytecode manipulation framework \cite{296} (as used in \cite{51}). The metrics calculated in the regenerated datasets and used in this study include complexity metrics (i.e., WMC \cite{134} and NOM \cite{147}) as well as coupling metrics (i.e., CBO \cite{121}, DCC - Direct Class Coupling \cite{133}, Export and Import coupling \cite{216}).


These metrics are selected to cover software quality aspects from different perspectives with varying focus. WMC captures the number of methods in a class weighted by their cyclomatic complexity. If weight is considered to be unity for all methods, it will effectively be equal to the number of methods, which is captured by NOM. CBO measures the number of other classes with which a class is coupled. Two classes are considered coupled if a method in one of them uses a method or an attribute defined by the other. The DCC of a class, however, counts the number of other classes to which the class is coupled by attribute declarations or parameter passing. Export and Import coupling capture the number of all other classes that reference and are referenced by the class (incoming and outgoing connections), respectively.






\subsection{Analysis and Testing}
\label{sec:MethodAnalysis}

\subsubsection{Threshold Calculation}

Univariate LR models are first fitted for each of the six metrics in all $36$ datasets, one by one. Whenever a significant association (at $p{<}0.05$) is found in a dataset between a metric and the defect probability, a threshold is calculated by using Eq.~\ref{eq:VARL} for the metric in the corresponding system. Doing so for all metrics in all systems creates six lists (one for each metric) of $36$ \textsl{size}-\textsl{threshold} pairs. A snippet of the list created for the CBO metric is shown as an example in Table~\ref{tab:LogisticPairs}, where $x$ indicates that CBO does not have a significant association with defects in the corresponding system.

\subsubsection{Correlation Analysis}

The correlation between sizes and thresholds (within the list of each metric) is then analyzed using the Spearman's rank correlation coefficient. However, because there are multiple \textsl{size}-\textsl{threshold} pairs representing multiple versions of the same systems (as can be seen in Table~\ref{tab:LogisticPairs}), only one pair from each system is included in the correlation analysis\footnote{The pairs selected are those representing the latest version of each system.}. This is done to avoid having a correlation that is due to factors other than system size (e.g., quality improvement in subsequent larger versions, which could enable the calculation of larger thresholds).


\subsubsection{Threshold Estimation Models}

\begin{table}
\begin{minipage}{\columnwidth}
\centering 
\caption{A Snippet of the Size-Threshold Pairs as Calculated by LR Threshold Models for CBO} 
\begin{tabular}{l c c} 
\toprule 
& \multicolumn{2}{c}{\textbf{CBO}} \\ 
\cmidrule(l){2-3} 
\textbf{System} & System Size & Threshold\textsuperscript{$\ast$} \\ 
\midrule 
Ant v1.3 & 116   & x \\
Ant v1.4 & 160   & x \\
Camel v1.0 & 195   & 3 \\
Camel v1.2 & 368   & 3 \\
Camel v1.4 & 473   & 4 \\
Camel v1.6 & 523   & 3 \\
Eclipse JDT Core v3.4 & 994   & 14 \\
Eclipse PDE UI v3.4.1 & 1450  & 16 \\
Ivy v1.1 & 111   & 3 \\
Ivy v1.4 & 241   & 10 \\
Ivy v2.0 & 352   & 9 \\
JEdit v3.2.1 &  163  & 3 \\
........ & ...   & .. \\
\bottomrule 
\multicolumn{3}{c}{} \\
\multicolumn{3}{c}{\textsuperscript{$\ast$}Thresholds are rounded to the nearest integer.}
\end{tabular}
\label{tab:LogisticPairs} 
\end{minipage}
\end{table}

For all metrics with a statistically significant (at $p{<}0.05$) size--threshold correlation, linear regression models can be fitted and used to estimate their thresholds in other unseen systems. These models will henceforth be referred to as threshold estimation models, and they relate the independent variable (system size) to the dependent variable (threshold). However, to assess the feasibility of this approach as a threshold estimation method, the performance of the estimation models needs to be appropriately tested. This is done by evaluating the accuracy with which fault-prone classes are identified using the estimated thresholds. The testing procedure followed is similar to, but slightly different from, the leave-one-out cross-validation (LOOCV), as explained in the following step.

\subsubsection{Model Testing}

In each round of a standard LOOCV procedure, one (test) data point would be removed and a model would be fitted on the $35$ remaining (training) data points. However, because there are multiple versions of the same systems, one system version may be used for testing while other versions of the same system are used for training. This is a problem because different versions of the same system are likely to have a significant proportion of the same classes, and many of these classes have similar metric values across different versions. For example, the class \texttt{org. apache.camel.Component} has almost identical metric values in all four Apache Camel versions. Therefore, to have totally independent training and testing sets, one test system is picked in each round of the LOOCV, and all of its versions are removed from the training set. When, for example, the Camel system is picked as a test system, all four versions of the system are removed. The model fitted on the remaining data points is then used to estimate test thresholds for each of the test versions. This process is repeated until each metric in every system version is assigned a test threshold. Table~\ref{tab:TestThresholds} shows a snippet of the test thresholds for CBO as an example. It is these thresholds that are used during the accuracy evaluation of the threshold estimation models.

\begin{table}[t]
\begin{minipage}{\columnwidth}
\centering 
\caption{A Snippet of the Test Thresholds as Estimated by Threshold-Estimation Models for CBO} 
\begin{tabular}{l c} 
\toprule 
& \multicolumn{1}{c}{\textbf{CBO}} \\ 
\cmidrule(l){2-2} 
\textbf{System} & Threshold\textsuperscript{$\ast$} \\ 
\midrule 
Ant v1.3   & 5 \\
Ant v1.4   & 5 \\
Camel v1.0   & 6 \\
Camel v1.2   & 8 \\
Camel v1.4   & 9 \\
Camel v1.6   & 9 \\
Eclipse JDT Core v3.4   & 13 \\
Eclipse PDE UI v3.4.1  & 19 \\
Ivy v1.1   & 5 \\
Ivy v1.4   & 6 \\
Ivy v2.0   & 7 \\
JEdit v3.2.1 & 5 \\
........ & ..  \\
\bottomrule 
\multicolumn{2}{c}{} \\
\multicolumn{2}{c}{\textsuperscript{$\ast$}Thresholds are rounded to the nearest integer.}
\end{tabular}
\label{tab:TestThresholds} 
\end{minipage}
\end{table}

\subsection{Evaluation}
\label{sec:MethodEvaluation}

The threshold estimation models, as well as the LR threshold models, are evaluated based on the accuracy with which their thresholds can classify classes into faulty and non-faulty categories. Classification accuracy is measured using the geometric mean (g-mean) which, unlike the f-measure, is insensitive to the imbalance in the datasets \cite{91}. The g-mean is calculated based on recall, which is the proportion of true positives to the total number of positives, and specificity, which is the proportion of true negatives to the total number of negatives. A software class with a metric value greater than or equal to a threshold is classified as faulty. Since all the metrics included in this study can only have discrete values, all thresholds are rounded to the nearest integer. Also, the minimum threshold for all metrics is set to two. Otherwise, any class that does anything or connect with any other class would be identified as a faulty class, leading to many false positives.

The threshold estimation models are evaluated using the test thresholds, as discussed in the preceding section. For example, the test threshold for CBO in the Eclipse JDT Core is $13$ (see Table~\ref{tab:TestThresholds}). Thus, any class with a CBO value of $13$ or more is classified as a (potentially) faulty class, and non-faulty otherwise. The accuracy of this classification is then measured using g-mean, and the same is done for each of the six metrics in all $36$ systems. The g-mean values achieved by the thresholds of the different metrics are used to analyze and compare the performance of the estimation models.

When evaluating the LR threshold models, on the other hand, the thresholds used are calculated after fitting the test datasets themselves. This gives an advantage to the LR threshold models compared with the threshold estimation models, as the former models have full exposure to the metric values and defect probabilities in the test systems, while the latter models get only the size of the test systems as input, based on which thresholds are estimated. Nevertheless, comparing their accuracy should provide important insight into the feasibility and efficacy of estimating thresholds given only system sizes.



%

The tests described above evaluate the accuracy of the thresholds of different metrics separately. To test their combined effect on classification accuracy, another test is conducted by training Naive Bayes (NB) models with and without thresholds (henceforth referred to as threshold and no-threshold NB models). The selection of the NB model is motivated by its good performance in similar studies \cite{145}. For each one of the $36$ datasets, NB models are constructed and tested using the LOOCV. Metrics are binarized for the threshold NB models using the test thresholds prepared for the estimation models as well as the LR model thresholds, resulting in two versions of these models. The binarized metric value for a class $c$ is calculated as follows:
\vspace{-1mm}
\begin{equation}
    Metric^*(c)\:= \begin{cases}
                1       	& if\; Metric(c) \geq threshold \\
                0               	& Otherwise
							\end{cases}
\end{equation}

This method of binarizing metric values is similar to the method used in \cite{291}. The classification accuracy of the threshold NB models is then compared with the accuracy of the no-threshold models as a baseline. The statistical test recommended in \cite{274} for comparing multiple classifiers (i.e., threshold and no-threshold NB models) over multiple~($36$) datasets is the Friedman non-parametric test. The null hypothesis to be tested is that all the classifiers are equivalent. If this null hypothesis is rejected at $p{<}0.05$, the Nemenyi post-hoc test is applied to make pairwise comparisons.



\section{Results and Discussion}
\label{sec:Results}

Table~\ref{tab:sigcounts} shows the number of systems in which metrics have a significant association with defects. As can be seen in the table, all metrics have a significant association in at least $30$ of the $36$ systems; the only exception is the Export coupling metric which has a significant association in only $15$ systems. The difference between Import and Export coupling metrics in their relationship with defects is consistent with the results reported in the literature \cite{105}.

\begin{table}[!t]
\renewcommand{\arraystretch}{1.26}
\begin{minipage}{\columnwidth}
  \centering
  \caption{Number of Systems in Which Metrics Have a Significant Association With Defects}
    \begin{tabular}{?c?c?}
		\Xhline{0.5pt}
    \multirow{2}[4]{*}{\textbf{Metric}} & \textbf{Significance Count} \\
\cline{2-2}          & (out of 36) \\
		\Xhline{0.5pt}
    CBO   & 30 \\
		\Xhline{0.25pt}
    DCC   & 30 \\
		\Xhline{0.25pt}
    Export Coupling & 15 \\
		\Xhline{0.25pt}
    Import Coupling & 33 \\
		\Xhline{0.25pt}
    NOM   & 34 \\
		\Xhline{0.25pt}
    WMC & 31 \\
		\Xhline{0.5pt}
    \end{tabular}%
  \label{tab:sigcounts}%
	\end{minipage}
\end{table}%

\begin{table*}[b]
\renewcommand{\arraystretch}{1.10}
  \centering
  \caption{Spearman's Rank Correlation Coefficients (and P-values) Before and After Applying the Correction}
    \begin{tabular}{?c?c|c?c|c?c|c?}
		\Xhline{0.5pt}
    \multirow{2}[4]{*}{Metric Name} & \multicolumn{2}{c?}{Risk Increase: +0.05} & \multicolumn{2}{c?}{Risk Increase: +0.10} & \multicolumn{2}{c?}{Risk Increase: +0.15} \\

\cline{2-7}          & Before & After & Before & After & Before & After \\

		\Xhline{0.5pt}
    \multirow{2}[4]{*}{CBO} & 0.45  & \textbf{0.63} & 0.5   & \textbf{0.63} & 0.5   & \textbf{0.63} \\
		
\cline{2-7}          & (0.147) & (0.032) & (0.099) & (0.032) & (0.099) & (0.032) \\

		\Xhline{0.5pt}
    \multirow{2}[4]{*}{DCC} & \textbf{0.69} & \textbf{0.87} & \textbf{0.76} & \textbf{0.88} & \textbf{0.76} & \textbf{0.88} \\
		
\cline{2-7}           & (0.011) & (0.000) & (0.004) & (0.000)   & (0.004) & (0.000) \\

		\Xhline{0.5pt}
    \multirow{2}[4]{*}{Export Coupling} & 0.55  & \textbf{0.95} & 0.62  & \textbf{0.95} & 0.71  & \textbf{0.95} \\
		
\cline{2-7}           & (0.171) & (0.001) & (0.115) & (0.001) & (0.058) & (0.001) \\

		\Xhline{0.5pt}
    \multirow{2}[4]{*}{Import Coupling} & 0.55  & \textbf{0.67} & 0.55  & \textbf{0.67} & 0.54  & \textbf{0.67} \\
		
\cline{2-7}           & (0.071) & (0.020) & (0.071) & (0.020) & (0.075) & (0.020) \\

		\Xhline{0.5pt}
    \multirow{2}[4]{*}{NOM} & 0.31  & \textbf{0.59} & 0.35  & \textbf{0.59} & 0.39  & \textbf{0.58} \\
		
\cline{2-7}           & (0.306) & (0.0381) & (0.239) & (0.038) & (0.189) & (0.042) \\

		\Xhline{0.5pt}
    \multirow{2}[4]{*}{WMC} & 0.18  & 0.42  & 0.18  & 0.43  & 0.24  & 0.43 \\
		
\cline{2-7}           & (0.566) & (0.152) & (0.566) & (0.140) & (0.425) & (0.140) \\

		\Xhline{0.5pt}
    \end{tabular}%
  \label{tab:correlations}%
\end{table*}%


\subsection{Size--Threshold Correlation}

Table~\ref{tab:correlations} shows the correlation coefficients for the size--threshold relationships, calculated before and after applying the correction (Eq.~\ref{eq:ch4correction}), at three different levels of acceptable increase to the background risk (Eq.~\ref{eq:ch4Parl}). It can be noted in the table that all metric thresholds (except those for WMC) have a significant positive correlation with software size, but only after the correction is applied. The absence of any correlation before, and the strong correlation after, has prompted us to examine the correlation between thresholds and the ratio of defects. Interestingly, a significant negative correlation is found in four (CBO, Import coupling, NOM and WMC) of the six metrics at all risk-increase levels. This may be explained by the fact that because data imbalance leads to underestimating risk, a lower defect ratio results in greater underestimation of risk and higher thresholds, and hence the negative correlation. So, without the correction, the calculated thresholds will be influenced by random fluctuations in defect ratios. This demonstrates the impact of data imbalance and supports the importance of considering its effect in similar studies.


Another interesting observation in Table~\ref{tab:correlations} is that most correlation coefficients are stable across different risk-increase levels (after the correction is applied). This indicates that the ranks of thresholds do not change despite the changes in threshold values. It is worth noting that all metric thresholds have a minimum of $10$ degrees of freedom, except Export coupling which has six degrees of freedom.

The thresholds used for the classification tests (discussed in the next section) are those calculated at the risk-increase level of $0.10$ because the correlations are slightly better at this level. Since WMC does not have any significant correlation with size, it is henceforth not discussed in this paper.

\begin{figure}[!t]
\centering
\includegraphics[width=\linewidth]{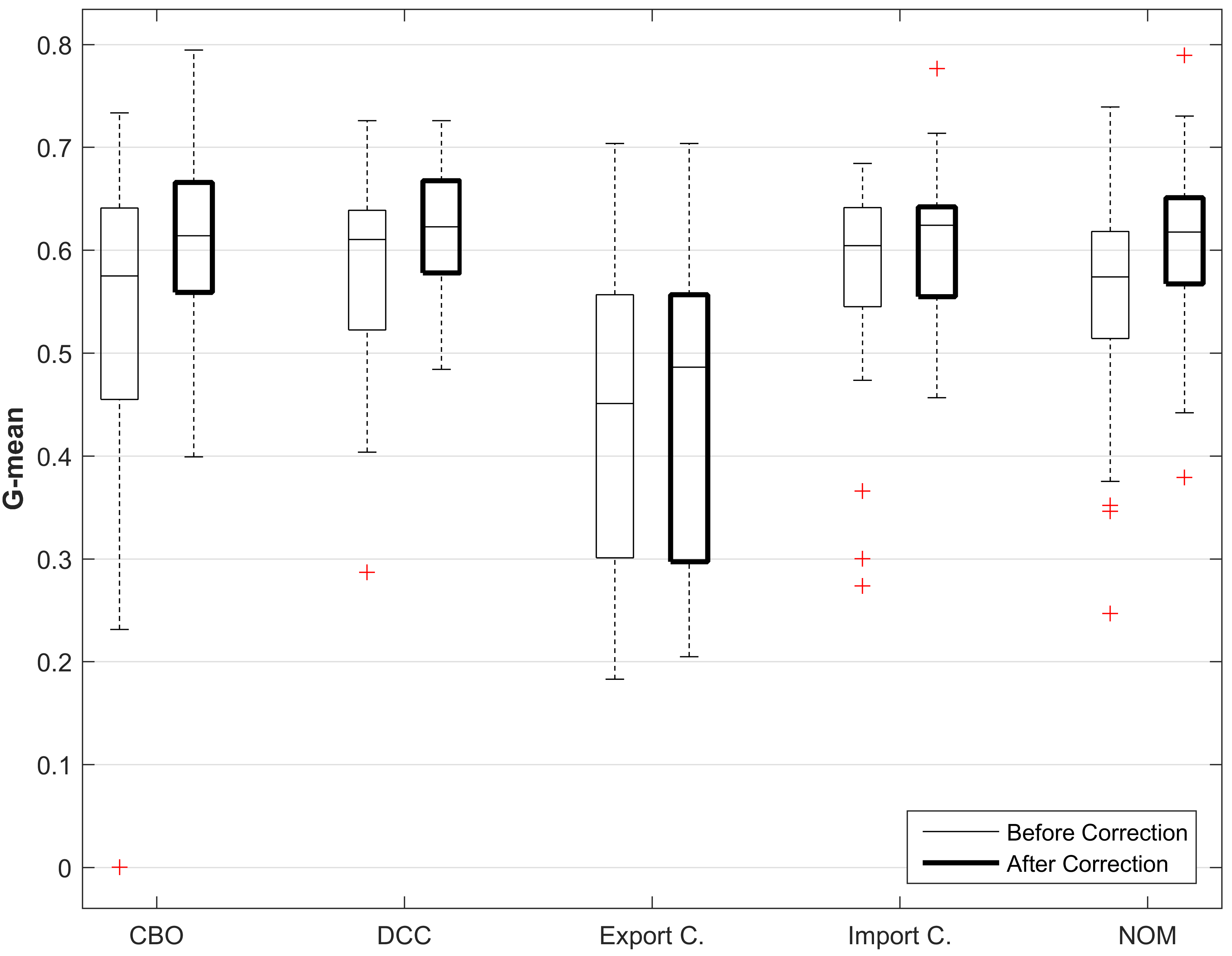}
\caption{Classification Accuracy for the LR Threshold Model Before and After Applying the Correction}
\label{LRBeforeAfter}
\end{figure}

\subsection{Classification Accuracy}

The impact of applying the correction on the classification accuracy of the LR model thresholds is shown in Fig.~\ref{LRBeforeAfter}. It is evident that the correction has a positive impact on the accuracy of all metric thresholds, as shown by the increase in the median g-mean for all metrics. The accuracy improvement is particularly evident in CBO, DCC and NOM. In addition, while all metric thresholds achieve a g-mean of at least 0.60 for the majority of the datasets (after correction), Export coupling is again an exception as its thresholds have a g-mean of less than $0.50$ in more than 50\% of the datasets. The relatively low accuracy of the Export coupling thresholds is consistent with the results reported in the literature, as mentioned earlier.

\begin{figure}[!t]
\centering
\includegraphics[width=\linewidth]{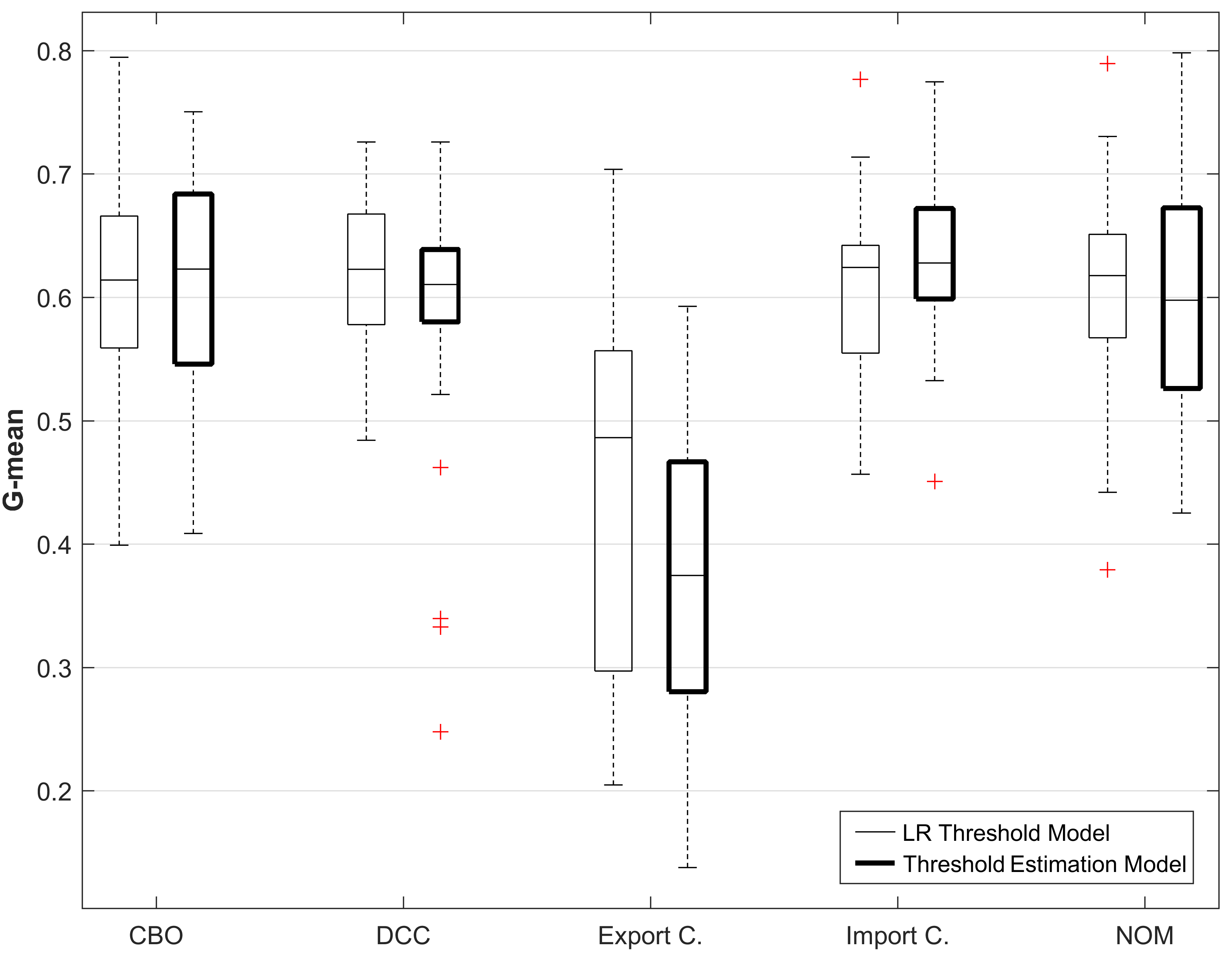}
\caption{Classification Accuracy for the LR Threshold Model and the Threshold Estimation Model}
\label{LogisticLinear}
\end{figure}

\begin{figure}[!t]
\centering
\includegraphics[width=\linewidth]{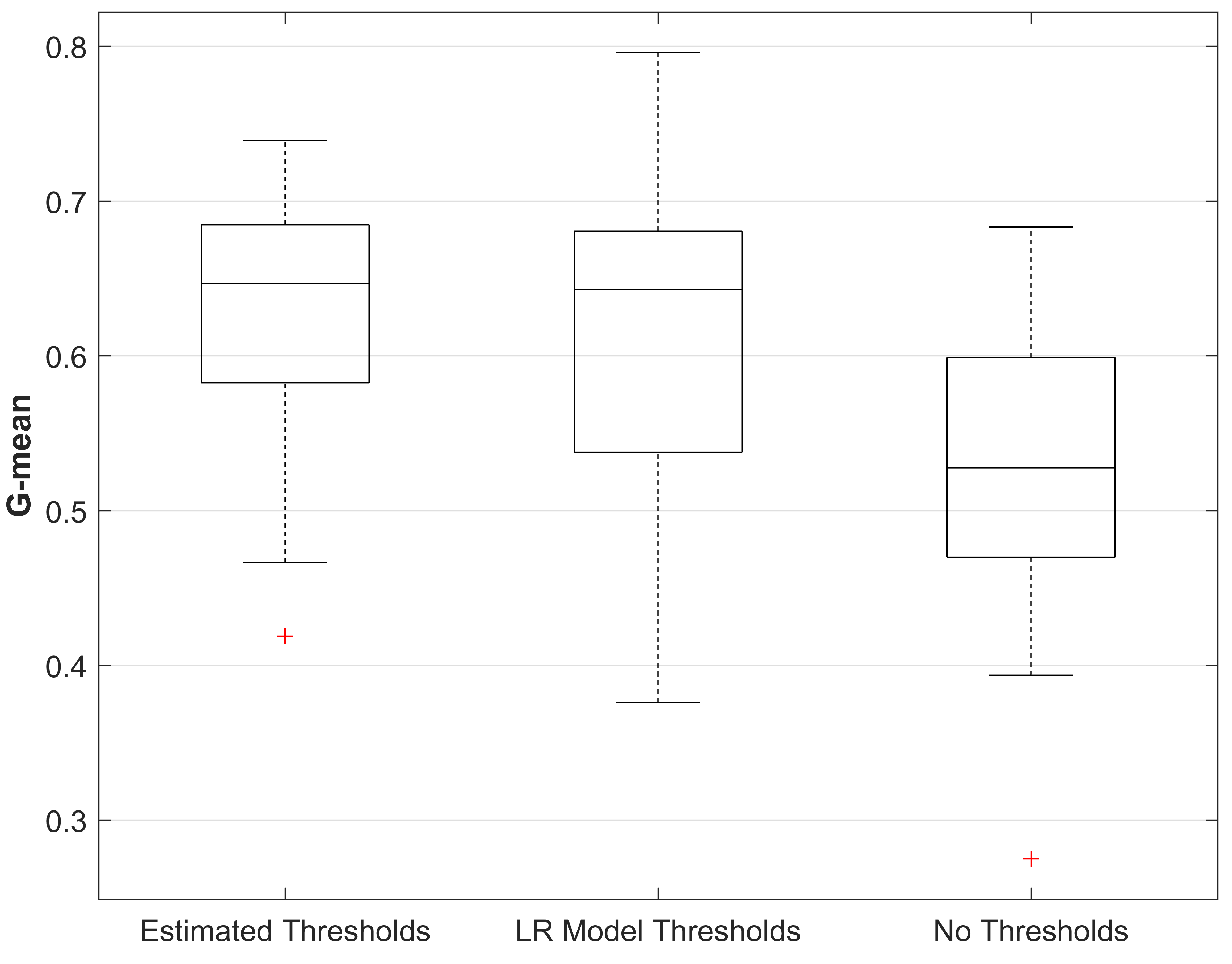}
\caption{Classification Accuracy for Threshold and No-Threshold NB Models}
\label{ThreeModels}
\end{figure}

Fig.~\ref{LogisticLinear} shows a comparison between the classification accuracy of the LR model thresholds (after correction) and the estimation model thresholds. The thresholds of both models have a similar overall accuracy for all metrics apart from the Import and Export coupling metrics; the estimation model has better accuracy in the former metric and worse in the latter one. The fact that Export Coupling has fewer data points to fit the estimation model on\footnote{Export Coupling has a significant association with defects in only $15$ systems, and thus there are relatively few \textsl{size}-\textsl{threshold} pairs to fit.} may explain its lower accuracy. Nevertheless, it is interesting that the estimation model has achieved an accuracy comparable to that of the LR threshold model, considering that the LR model has a full exposure to the metric values and defect probabilities in the test systems whereas the estimation model does not, as explained earlier. The same can also be said in comparison with the results reported by other studies (e.g. \cite{297} and \cite{298}) with threshold models that are constructed using more than just size-\textsl{threshold} pairs.



\begin{figure}[!b]
\centering
\includegraphics[width=\linewidth]{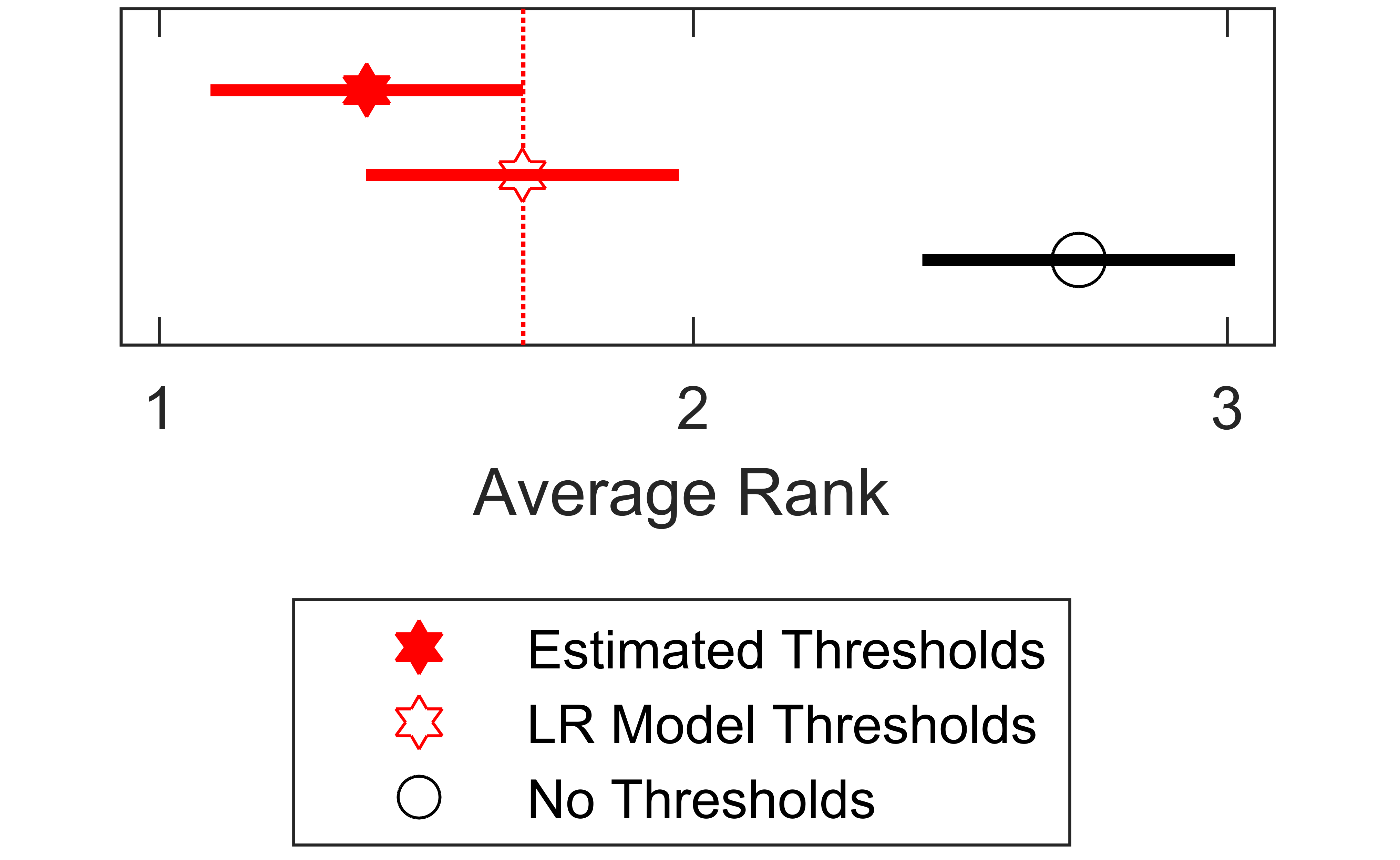}
\caption{Nemenyi Post-Hoc Test Result for Threshold and No-Threshold NB Models}
\label{NemenyiPostHoc}
\end{figure}

For the purpose of evaluating the effect of all the thresholds together on the classification accuracy, NB models are trained and tested on the $36$ datasets as prescribed in Section~\ref{sec:MethodEvaluation}. Fig.~\ref{ThreeModels} shows the classification accuracy achieved by the two thresholds models as well as the no-threshold model. It can be seen that there is no apparent difference in the accuracy of the two threshold versions of the NB model, but they both appear to be more accurate than the no-threshold NB model. Using the Friedman test\footnote{The test actually used is Skillings–Mack \cite{275}, which is equivalent to the Friedman test but used in cases where there are missing values. The missing values in our case correspond mainly to systems with no LR thresholds (i.e., the LR version of the threshold NB models is missing for such systems).} shows that there is a statistically significant (at $p{<}0.05$) difference between the ranks of the three NB models. Therefore, the Nemenyi post-hoc test is applied to make pairwise comparisons; the results of which are shown in Fig.~\ref{NemenyiPostHoc}. The length of the horizontal lines represents the critical value, and the average ranks of any two classification models must be different by at least the critical value for them to be considered significantly different. It can be clearly seen that there is indeed no significant difference between the accuracy of the threshold NB models. A significant difference, however, exists between both threshold models and the no-threshold model. The threshold models have lower average ranks, and hence better accuracy.


The slopes and intercepts of the threshold estimation models that are fitted on all of the available \textsl{size}-\textsl{threshold} pairs are presented in Table~\ref{tab:SlopIntercept}. These equations can easily be used to calculate relative thresholds for the five metrics in any open-source Java system.


\begin{table}[!t]
\renewcommand{\arraystretch}{2}
\begin{minipage}{\columnwidth}
  \centering
  \caption{The Slopes and Intercepts of the Thresholds Estimation Models}
    \begin{tabular}{?c?c|c?}
		\Xhline{0.5pt}
    \textbf{Metric} & \textbf{Slope} & \textbf{Intercept} \\
		\Xhline{0.25pt}
    \textbf{CBO} & 0.00958 & 3.69029 \\
		\Xhline{0.25pt}
    \textbf{DCC} & 0.00432 & 0.50917 \\
		\Xhline{0.25pt}
    \textbf{Export Coupling} & 0.02162 & 2.51718 \\
		\Xhline{0.25pt}
    \textbf{Import Coupling} & 0.00476 & 2.07069 \\
		\Xhline{0.25pt}
    \textbf{NOM} & 0.00810 & 4.98745 \\
		\Xhline{0.5pt}
    \end{tabular}%
  \label{tab:SlopIntercept}%
\end{minipage}
\end{table}%

\section{Threats to Validity}
\label{sec:ValidityThreats}

\subsection{Construct Validity}

Metric values often differ depending on the tool used to calculate them, and metric values calculated using different tools may deviate differently from their true values. This is why the datasets included in this study are regenerated using a single tool (the regenerated datasets are available online), as discussed in Section~\ref{sec:Method}. Although the tool used has been extensively tested, it still cannot be guaranteed that the values it calculates for the metrics always match their true values perfectly. This may, in turn, leads to variations in the calculated thresholds.

\subsection{Internal Validity}

In Section~\ref{sec:Results}, size-threshold correlations are analyzed at three different risk-increase levels, and only one of which is considered in the classification accuracy tests. Choosing different risk-increase levels may affect the results reported for the correlation analysis as well as the accuracy tests. Moreover, in order to apply the correction discussed in Section \ref{sec:Effect}, the average ratio of defects in the $36$ datasets included in this work is assumed to be the ratio of defects in the population of open-source Java systems. Applying the correction with a different population ratio of defects may also affect the results.

\subsection{External Validity}

Only $36$ datasets, representing different versions of $13$ open source Java systems, are included in this study. The selected datasets are limited to those with publicly available source code as it is required to regenerate the datasets. Thus, the results may not generalize to all software systems, especially those implemented in different contexts (i.e., proprietary software systems and systems developed using other programming languages).


 \section{Related Work}
\label{sec:Related}

In \cite{108}, an empirical study was conducted to test if OO metrics exhibit a threshold effect, meaning that the fault-proneness of a class remains constant until it reaches a threshold, after which it increases rapidly. While this was assumed to occur when the capacity limit of human short-term memory is reached, their results did not support the existence of such an effect. However, the definition of \textsl{threshold} adopted in this study does not assume such an effect.

The receiver operating characteristic (ROC) analysis was applied in \cite{111} to three releases of the same system in order to calculate thresholds for 12 metrics. The goal was to identify thresholds that classify classes into two categories (faulty versus non-faulty) as well as into three categories based on fault severity. However, the analysis failed to identify thresholds for the binary classification, and thresholds were identified for only 5 of the 12 metrics in two of the three categories.


Thresholds identified based on the experience of NASA and its software assurance technology center were presented in \cite{128}. It was, however, acknowledged that there was little application data support for the thresholds. Several experience-based thresholds were also proposed in \cite{147}. In \cite{142}, an approach was proposed to calculate thresholds based on mean and standard deviation. However, this approach assumes that software metrics are normally distributed, whereas, in fact, many of them are not \cite{120}.

A methodology for deriving thresholds from a large dataset was proposed in \cite{116}, in which the quantile function was used to identify threshold values that represent the majority (e.g. 80\%) of the dataset. The thresholds could then be used to identify the minority (e.g. 20\%) which has relatively extreme values. A similar approach was proposed in \cite{143} with the explicit aim of identifying relative thresholds, the value of which are larger than a specified minimum percentage of classes in a given system. The approach proposed in \cite{113} was based on fitting datasets to Poisson and Weibull probability distributions, which were then used to identify typical metric values (i.e. thresholds). Similarly, the Power-Law probability distribution function was used in \cite{126} to fit a dataset and calculate thresholds. Nevertheless, the results reported in \cite{289} suggest that distribution-based thresholds are unreliable compared with those obtained using methods that consider quality-related (e.g., defect) data.


\section{Conclusion and Future Work}
\label{sec:Conclusion}

The work presented in this paper assessed the feasibility of estimating thresholds for OO metrics based on, and relative~to, system size. To this end, thresholds were calculated for six metrics in 36 software systems of various sizes and the size--threshold relationship was investigated. Predictive models were then built to estimate thresholds, and the accuracy of the estimated thresholds in identifying fault-prone classes was evaluated. The results show that the proposed threshold estimation method is feasible, and that the predictive models can perform at an accuracy comparable to more complex threshold models with access to metric values and defect probabilities. 

The proposed method has two main advantages. First, estimation models, such as the ones built in this study, can be easily used to calculate relative thresholds, even in systems that lack historical quality-related data. Second, such models can be used both to identify faulty classes during software maintenance and to formulate relative guidelines for the development process.

An improved version of the LR threshold model was also proposed in this paper. The improvements aim to solve the problem of generating negative thresholds as well as to counteract the effect of data imbalance on the LR model. The~results show that the improved model's thresholds achieved a noticeably higher accuracy in identifying fault-prone classes than those of the original threshold model.

Although the overall results are encouraging, this study is more exploratory than conclusive. It is therefore suggested that the proposed threshold estimation method should be investigated further using more datasets, different metrics and different programming languages. Also, for future work, models other than the LR threshold model may be used to calculate thresholds in individual systems. Different types of estimation models may also be used to fit the size--threshold relationship.

\bibliographystyle{IEEEtran}
\bibliography{C:/Users/Sultan/Dropbox/00REFERENCES/main_references_list}

\end{document}